\documentstyle[prl,aps,epsf]{revtex}
\begin{document}
\draft
\twocolumn[\hsize\textwidth\columnwidth\hsize\csname
@twocolumnfalse\endcsname
\title{
Nonequilibrium Dynamic Phase Diagram for Vortex Lattices 
}
\author{C. J. Olson, C. Reichhardt, and Franco Nori}
\address{Department of Physics, The University of Michigan, Ann Arbor,
Michigan 48109-1120}
\date{\today}
\maketitle
\begin{abstract}
The new dynamic phase diagram for driven vortices with varying lattice softness
we present here indicates that,
at high driving currents, at least {\it two distinct dynamic phases} 
of flux flow appear depending on the vortex-vortex interaction strength.  
When the flux lattice
is soft, the vortices flow in independently moving channels with smectic
structure. 
For stiff flux lattices, 
adjacent channels become locked
together, producing crystalline-like order in a coupled channel phase.
At the crossover lattice softness between these phases, the system
produces a maximum amount of voltage noise.
Our results relate spatial order with transport and
are in agreement with experiments.
\end{abstract}
\vspace{-5pt}
\vspace{-15pt}
\vskip2pc]
\vskip2pc

Nonequilibrium problems involving elastic lattices and disordered media,
such as the nature of depinning transitions or the behavior of 
rapidly driven lattices, appear in a wide variety of systems including
superconducting vortex lattices, charge-density waves, and
solid-on-solid friction.
Much recent interest has been devoted to the motion of a vortex lattice (VL)
across a disordered substrate under applied currents
both at and well beyond depinning.
The reordering of a rapidly-driven VL is supported by 
simulations \cite{Shi,Koshelev,Moon,Spencer-Ryu,Reichhardt}
as well as 
neutron scattering \cite{Yaron} and
decoration experiments \cite{Duarte-Pardo-Marchevsky},
but the nature of the order 
that appears remains a subject of debate.  In addition,
the relationship of the voltage noise observed near depinning
to the microscopic vortex motion at higher currents has not been addressed.

Recent work on the behavior of a VL driven by a large
current produced conflicting pictures of the VL order, 
ranging from a crystalline 
structure \cite{Koshelev} 
to a moving smectic state \cite{Balents}.  
In Ref.~\cite{GiamarchiPRL94},
the VL does not fully recrystallize, but instead enters
a moving Bragg glass state with channels.
Moreover, Ref. \cite{Balents} describes the VL in terms of
independently moving channels of vortices with overall smectic order.
Other theories also focus on channels of vortices \cite{Scheidl}.
Smectic structure factors ($S({\bf q})$) of the VL
were observed in simulations of vortices moving
over strong pinning \cite{Moon,Spencer-Ryu}, in agreement with
Ref. \cite{Balents}.

Very recent decoration experiments \cite{NewPardo} produced {\it both\/}
crystalline-like
and smectic order of the moving VL, 
depending on the magnitude of the applied magnetic field.
Smectic order appears at low fields, when the vortices interact weakly,
and crystalline-like peaks in $S({\bf q})$ appear 
at high fields, when the vortex interactions are stronger.
This suggests that the softness of the VL is important
in determining the vortex behavior at high driving currents.

In this paper, we propose a new dynamic phase diagram 
in which {\it both\/}
smectic 
and crystalline-like order
appear as the VL softness
is varied.
Using simulations of current-driven vortices, 
we 
clearly 
define regions of the phase diagram
based on $S({\bf q})$,
$V(I)$ curves, voltage noise, velocity distributions,
direct observation of the
lattice, and defect density calculations.
We compute experimentally relevant
voltage noise spectra \cite{DAnna}
at all currents from depinning to high drives,
and find evidence for a washboard frequency in the stiff VL
as well as broad-band noise in the plastic flow phase.
As the VL softens, 
we observe a novel crossover to a regime in which 
smectic order is never destroyed even at high drives.
At the crossover, the 
amount of 
voltage noise generated by the system
during depinning is maximum,
indicating that experimental voltage measurements can 
probe the VL order.
Our results are in excellent agreement with experiments, and have
implications for noise measurements in the peak effect regime
\cite{Marley}.

We model a transverse two-dimensional slice (in
the $x$--$y$ plane) of a $T=0$ superconducting infinite slab containing
rigid vortices that are parallel to the sample edge 
(${\bf H}=H{\hat {\bf z}}$).  
The vortex-vortex repulsion is correctly represented
by a modified Bessel function, $K_{1}(r/\lambda)$,
where $\lambda$ is the penetration depth.
The vortices are driven through a sample with periodic boundary conditions,
filled with randomly placed columnar defects,
by a uniform Lorentz force  ${\rm f}_{d}{\hat {\bf x}}$, representing
an applied current.
The columnar pins are non-overlapping, short-range parabolic wells of 
radius $\xi_{p}=0.30\lambda$.  
The maximum pinning force, $f_{p}$, of wells 
has
a Gaussian distribution with a mean value of $f_{p} = 1.5 f_{0}$
and a standard deviation of $0.1 f_{0}$,
where $f_{0} = \Phi_{0}^{2}/8\pi^{2}\lambda^{3}$.  
The pin density $n_{p}=1.0/\lambda^{2}$ is higher than the
vortex density $n_{v}=0.7 \Phi_0/\lambda^{2}$.
We concentrate on samples $36\lambda \times 36\lambda$ in size,
containing 864 vortices and 1295 pins.
We check for finite size effects using samples
that range in size from $18\lambda \times 18\lambda$ to
$72\lambda \times 72\lambda$ and contain up to 
2484 vortices and 3600 pins.  

The overdamped equation
of vortex motion is
${\bf f}_{i}={\bf f}_{i}^{vv}+{\bf f}_{i}^{vp}+{\bf f}_{d}=\eta{\bf v}_{i} \ ,$
where the total force ${\bf f}_{i}$ on vortex $i$ (due to other vortices
${\bf f}_{i}^{vv}$, pinning sites ${\bf f}_{i}^{vp}$, and the driving
current ${\bf f}_{d}$) is
given by
${\bf f}_{i} = \ \sum_{j=1}^{N_{v}}\, A_{v} f_{0} \,\ K_{1} \hspace{-2pt}
\left( |{\bf r}_{i} - {\bf r}_{j} | /\lambda \right)
\, {\bf {\hat r}}_{ij}
+  \sum_{k=1}^{N_{p}} (f_{p}/\xi_{p}) \
|{\bf r}_{i} - {\bf r}_{k}^{(p)}| \ \ \Theta \hspace{-2pt} \left[
(\xi_{p} - |{\bf r}_{i} - {\bf r}_{k}^{(p)} |)/\lambda \right] \
{\bf {\hat r}}_{ik} + {\bf f}_{d}$.
Here, $ \Theta$ is the Heaviside step function,
${\bf r}_{i}$ (${\bf v}_{i}$) is the location (velocity) of the $i$th vortex,
${\bf r}_{k}^{(p)}$ is the location of the $k$th pinning site,
$\xi_{p}$ is the pinning site radius,
$N_{p}$ ($N_{v}$) is the number of pinning sites (vortices),
${\bf {\hat r}}_{ij}=({\bf r}_{i}-{\bf r}_{j} )/
|{\bf r}_{i} -{\bf r}_{j}|$,
${\bf {\hat r}}_{ik}=({\bf r}_{i}-{\bf r}_{k}^{(p)})/
|{\bf r}_{i}-{\bf r}^{(p)}|$.

Using the monotonic relationship between the shear modulus 
$c_{66}$ and the dimensionless
prefactor $A_{v}$ of the vortex-vortex interaction term,
we model VLs with varying degrees of softness by changing
$A_{v}$ over 3.5
{\it orders of magnitude\/}, 
from $A_{v} = 0.001$ to $A_{v} = 6.0$.
This is in contrast to previous simulations
\cite{Moon,Spencer-Ryu,Reichhardt,Jensen,Olson} 
that considered only 
an extremely {\it soft\/} VL.
For each value of $A_{v}$, we simulate a voltage-current $V(I)$ curve 
by initially placing the vortices in a perfect lattice.
slowly increasing the driving force ${\rm f}_{d}$ 
from zero to $3.0f_{0}$ and measuring the
resulting voltage signal $V_x = \sum {\rm f}_{x}^{(i)}/N_{v}$.  

In order to identify the phase boundaries, 
each time the driving current ${\rm f}_{d}$ increases by $0.08f_{0}$, 
we halt the increases in ${\rm f}_{d}$, check that
the voltage signal $V_{x}$ is stationary over time,
and then collect detailed velocity and position information for
a long time interval, $2 \times 10^{5}$ molecular dynamics (MD) steps,
at a single current.
For a $36 \lambda \times 36 \lambda$ sample containing 864 vortices,
each $V(I)$ curve requires $8 \times 10^{6}$ to $10^{7}$ MD steps
to complete, corresponding to about 10 days of CPU time on a SPARC Ultra.
At each current, we are able to compute accurate voltage noise
spectra $S(\omega)$ with a frequency window extending into relatively 
low frequencies on the order of the vortex time-of-flight across the sample.
We also determine the voltage noise power $S_{0}$ in one frequency octave, 
$S_{0} = \int_{\omega_0}^{\omega_1} d\omega S(\omega)$,
where $S(\omega) = | \int V_{x}(t) e^{-i2\pi\omega t}dt |^{2}$.
The units of frequency are inverse MD time steps.
Here, $\omega_0 = 0.027$ and $\omega_1 = 0.054$ were chosen to fall in
the middle of our frequency window.  The results are not affected if
nearby values are used.
To measure the order
in the VL, 
we use the Delaunay triangulation of the
instantaneous vortex positions to determine the fraction of 
six-fold coordinated vortices, $P_{6}$, and 
we compute the structure 
factor, 
$S({\bf q}) = |\sum_{i=1}^{N_{v}} e^{i{\bf q} \cdot {\bf r}_{i}}|^{2}/N_{v}$,
of the VL. 

\begin{figure}
\centerline{
\epsfxsize=3.5in
\epsfbox{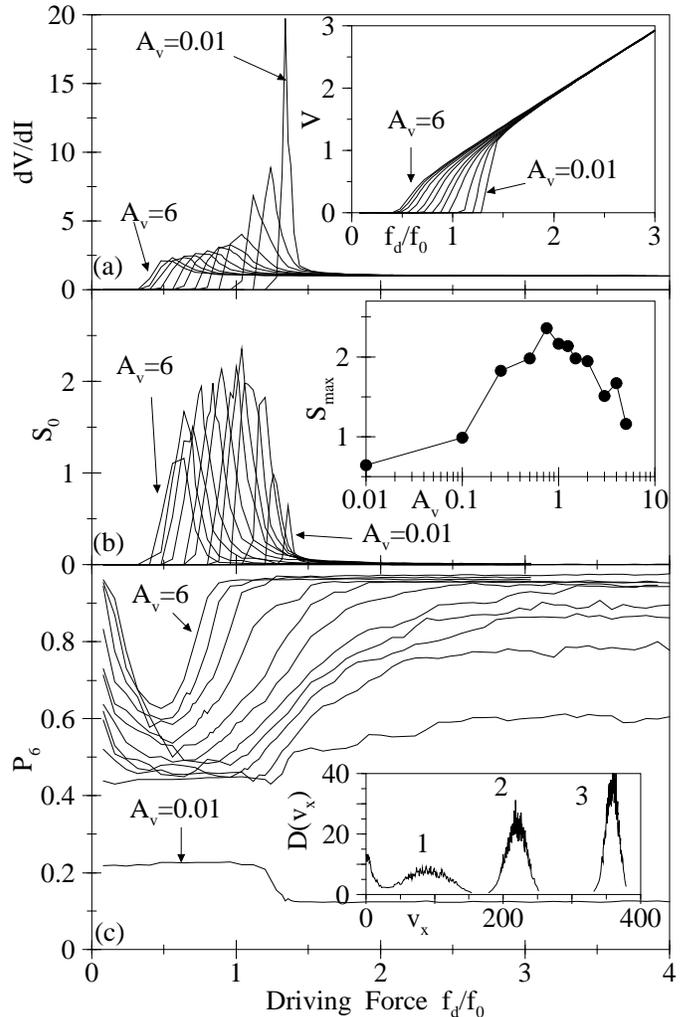}}
\caption{
(a): $dV/dI$ curves for vortex-vortex interactions
(right to left) $A_{v} = $ 0.01, 0.1, 0.25, 0.5, 0.75, 1, 1.5, 2, 3, 4, 5, 6.
The peak in $dV/dI$ increases in magnitude with decreasing $A_{v}$.
Inset to (a): Voltage-current $V(I)$ curves for the $A_{v}$ values listed
above.
The depinning transition shifts to higher driving
currents ${\rm f}_{d}$ and becomes more abrupt as $A_{v}$ decreases.
(b): Voltage noise power $S_{0}$ versus ${\rm f}_{d}$ for the $A_{v}$
values listed above.  In each case the noise power {\it peaks\/} 
during the plastic 
flow regime.  Inset to (b): Maximum noise power $S_{\rm max}$ as a function
of $A_{v}$.  The peak value of $S_{\rm max}$ corresponds to 
$A_{v} \approx 0.75$, the {\it same\/} $A_{v}$ at which the system crosses 
between smectic and coupled channel behavior at high drives.
(c): Fraction of six-fold coordinated vortices $P_{6}$ as a function
of ${\rm f}_{d}$ for values of $A_{v}$ listed above.
At zero drive and strong VL coupling, 
$P_{6} \sim 1$ since the VL is field-cooled.
The lowest value of $P_{6}$ at each $A_{v}$ corresponds {\it exactly\/} to the
peak in $dV/dI$.  The VL eventually reorders to $P_{6} \sim 1$ only
when $A_{v} \geq 1$.  For $A_{v} \leq 0.75$,
$P_{6}$ saturates at a value below 1.
Inset to (c): Velocity distribution functions $D(v_x)$
for (1) plastic, (2) smectic,
and (3) coupled channel phases.
}
\label{fig:triple}
\end{figure}

The depinning of the VL appears in our simulated
$V(I)$ curves, shown in the
inset to Fig.~\ref{fig:triple}(a).  
Since a soft VL is able to deform and take better advantage of
the available pinning sites, as $A_{v}$ decreases the depinning transition 
shifts to higher driving forces ${\rm f}_{d}$, 
from 
${\rm f}_{d}=0.48f_{0}$ at $A_{v}=6.0$ to 
${\rm f}_{d}=1.36f_{0}$ at $A_{v}=0.01$.
The depinning also becomes more abrupt for softer VLs,
producing a peak in $dV/dI$ that grows in magnitude with
lower $A_{v}$, as seen in Fig.~\ref{fig:triple}(a).  
The fact that a {\it soft VL has a higher depinning current\/} 
agrees well with
experiments \cite{NewPardo,Marley}.  

The VL also produces the greatest voltage noise at currents
{\it just above depinning\/}, when 
the vortices are in a plastic flow regime.
As seen in Fig.~\ref{fig:triple}(b), for each
value of $A_{v}$
the noise power $S_0$ reaches its peak value $S_{\rm max}$ at a driving
current ${\rm f}_{d}$ {\it above\/} the current at which 
the peak 
in $dV/dI$
occurs.  The current ${\rm f}_{d}$ at which the 
peak occurs is independent of sample size.
The spectra $S(\omega)$ 
near $S_{\rm max}$ are of the form
$S(\omega) \sim 1/\omega$, as illustrated in Fig.~\ref{fig:phase}(A).
As $A_{v}$ decreases and the VL softens, we 
observe a peak 

\hspace{-13pt}
in $S_{\rm max}$ at $A_{v} \sim 0.75$, shown in the inset 
to Fig.~\ref{fig:triple}(b). It is important to point out that this 
result 
agrees well with experiments conducted in the peak effect regime 
\cite{Marley}, 
in which an observed peak in voltage noise power 

\begin{figure}
\centerline{
\epsfxsize=3.5in
\epsfbox{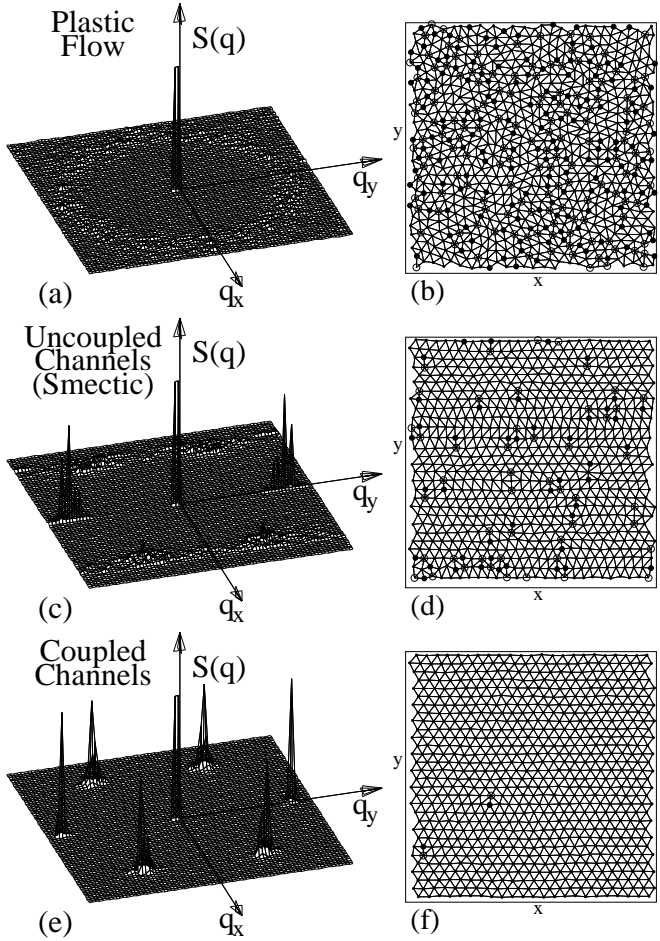}}
\caption{Structure factor $S({\bf q})$ 
(a,c,e) and Delaunay triangulations (b,d,f)
of the VL for: (a,b) $A_{v} = 1.5$, ${\rm f}_{d} = 1.12f_{0}$, 
in the
plastic flow regime.  $S({\bf q})$ 
is liquid-like, and the VL is filled with
defects (marked with circles).  (c,d) $A_{v} = 0.5$, 
${\rm f}_{d} = 3.04f_{0}$, in the uncoupled channel
regime.  $S({\bf q})$ has smectic peaks, and the defects in the VL are
oriented transverse to the $x$--direction driving force.  (e,f)
$A_{v}=4$, ${\rm f}_{d}=3.04f_{0}$, in the coupled channel regime.  
$S({\bf q})$ has
slightly anisotropic crystalline-like peaks, and the VL contains
almost no defects.
}
\label{fig:sq}
\end{figure}

\hspace{-13pt}
with changing magnetic field is interpreted to occur due to
the softening of the VL as the externally applied magnetic 
field increases.

We quantify changes in the VL
structure by calculating the fraction of six-fold coordinated
vortices $P_{6}$, shown in Fig.~\ref{fig:triple}(c), and
the structure factor $S({\bf q})$, shown in Fig.~\ref{fig:sq}.
The vortices are initially field-cooled,
so 
$P_{6} \approx 1$ at ${\rm f}_{d} = 0$.  
At driving currents below the depinning transition, the VL relaxes
into the pinning sites, causing a gradual decrease in $P_{6}$.
The relaxation is
most pronounced for soft VLs with low values of $A_{v}$.
A plastic flow state appears just above depinning, producing a liquid-like
structure factor $S({\bf q})$, shown in Fig.~\ref{fig:sq}(a).
The Delaunay triangulation in Fig.~\ref{fig:sq}(b) reveals the large number
of defects in the VL in this regime.
As shown in Fig.~\ref{fig:triple},
we find that the peak in $dV/dI$ corresponds {\it exactly\/} to the driving 
force at which $P_{6}$ reaches its {\it lowest\/} value.
Thus $dV/dI$ can be used to probe the order in the VL.

As seen by the rise of $P_{6}$ in Fig.~\ref{fig:triple}(c)
for $A_{v} \ge 0.1$,
the VL regains order as the driving force ${\rm f}_{d}$ is
increased and the interactions with pinning sites become less important.
In stiff VLs with large values of $A_{v}$, 
the defects quickly heal out and $P_{6}$ returns to a value near 1 for
driving forces ${\rm f}_{d}$ not much larger than the pinning force $f_{p}$.
As $A_{v}$ is lowered and
the VL softens, higher and higher
drives ${\rm f}_{d}$ must be applied to bring $P_6$ back to 1.
In Fig.~\ref{fig:triple}(c), for $A_{v} \lesssim 0.75$,
we see that $P_{6}$ saturates at a value less than 1, and
does not increase further even after applying
driving forces ${\rm f}_{d}$ considerably
larger than those shown in the figure $({\rm f}_{d} \sim 6f_{0})$.
The {\it permanent presence of a significant number of defects\/}
represents an important change caused by the VL softness
in the behavior of the system at high driving currents.

Computations of the structure factor $S({\bf q})$, shown in Fig.~\ref{fig:sq},
verify that two different types of VL order appear
at high driving currents depending on the lattice softness.
When $A_{v}$ is low and the VL is correspondingly soft, 
the structure factor 
$S({\bf q})$ [Fig.~\ref{fig:sq}(c)] has a smectic appearance, 
with  well-defined peaks only along the $q_x = 0$ axis.
A Delaunay triangulation of the moving VL, shown in Fig.~\ref{fig:sq}(d),
reveals that the vortices are flowing in channels oriented in 
the $x$--direction (parallel to ${\rm f}_{d}$) and approximately regularly 
spaced in the $y$--direction.
(This is in contrast to the plastic flow regime, in which vortices that
remain pinned act as barriers and result in vortex
motion transverse to the direction of flow, preventing the formation of 
longitudinal, straight channels.)
Vortices in adjacent channels
interact so weakly that each channel is able to move independently of
nearby channels.  
All of the defects remaining in the VL
have their Burgers vectors aligned {\it transverse\/} to ${\rm f}_{d}$, 
and each channel passes through at least one defect.  Thus the channels
can easily slip past each other as the defects glide, resulting
in an uncoupled channel phase.

If $A_{v}$ is large and the VL is correspondingly stiff,
$S({\bf q})$ has a crystalline-like structure,
shown in Fig.~\ref{fig:sq}(e), at high driving currents.
The peaks in $S({\bf q})$ are anisotropic, with slightly higher peaks at 
$q_{x} = 0$ than at $q_{x} \ne 0$.
For larger samples, the same behavior is observed in $S({\bf q})$
except the peaks are better resolved.
The VL is nearly defect-free, as seen in the Delaunay triangulation
of Fig.~\ref{fig:sq}(f).  The vortices still move in channels oriented in
the direction of drive, but these channels are now locked together
by the strong vortex-vortex interactions.
Thus, the system is in a coupled channel dynamic state.

We summarize the transitions between different states of the
moving VL when $A_{v}$ and ${\rm f}_{d}$ are varied in the phase diagram 
shown in Fig.~\ref{fig:phase}.  
The boundary between the pinned and plastic flow phases reflects the
increase in depinning current as $A_{v}$ decreases, noted earlier.
The 
plastic flow regime, identified by its liquid-like
$S({\bf q})$ 
and by its bimodal vortex velocity distribution 
[inset to Fig.~\ref{fig:triple}(c)],
disappears above ${\rm f}_{d} \sim f_{p}$.
For ${\rm f}_{d} > f_{p}$, 
the softness of 

\begin{figure}
\centerline{
\epsfxsize=3.5in
\epsfbox{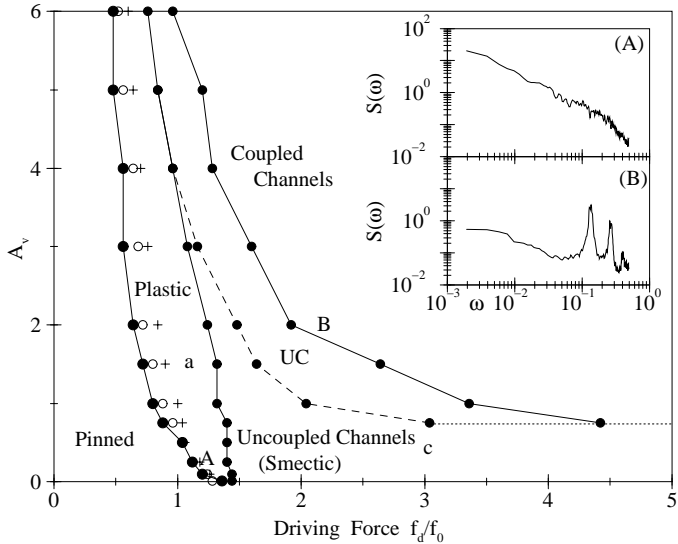}}
\caption{Dynamic phase diagram for different driving currents ${\rm f}_{d}$
and vortex-vortex interaction strengths $A_{v}$.  
At very low drives, the VL is pinned.
In the plastic flow phase, we observe {\it peaks\/} in $dV/dI$ (crosses) as
well as {\it peaks\/} in the voltage noise power $S_{0}$ (circles).
When $A_{v} \leq 0.75$, the VL flows in 
uncoupled channels for {\it all\/} 
high driving currents.  For larger $A_{v}$, the
VL passes through a transition region (UC) in which the channels
progressively couple, until reaching a reordered state at
high drives.  
A: Typical voltage noise spectrum
$S(\omega) \sim 1/\omega$ observed near $S_{\rm max}$ for $A_{v} = 0.1$.
B: Washboard frequency observed in the coupled channel
regime for $A_{v} = 2$.  The magnitude of the narrow band signal
decreases in samples larger than the $36\lambda \times 36\lambda$
sample shown here.  The letters A,B refer to inset plots; a,c refer
to Fig.~\ref{fig:sq} plots.}
\label{fig:phase}
\end{figure}

\hspace{-13pt}
the VL determines the behavior
of the system.  When $A_{v} \lesssim 0.75$, 
the structure factor has a smectic appearance 
for {\it all\/} drives ${\rm f}_{d} \gtrsim f_{p}$, 
corresponding to an {\it uncoupled\/} channel (UC)
flow regime.  
For $1 \lesssim A_{v} \lesssim 3$, the system enters this same smectic
UC state immediately after 
leaving the plastic flow state.
At slightly higher driving 
currents, however, the defect-filled VL enters
a transition regime marked by the appearance of weak peaks in 
$S({\bf q})$ at $q_{x} \ne 0$.  
These peaks are significantly 
smaller than the smectic peaks at $q_{x} = 0$, but are at least twice as 
large as the background observed in the smectic phase.
Throughout the UC transition regime, the channels gradually couple
as the number of defects in the VL drops.
The unimodal vortex velocity distribution becomes less broad, as in 
the inset to Fig.~\ref{fig:triple}(c).
When $A_{v} \gtrsim 4$, the system goes directly into this transition state
without ever passing through a truly smectic state.
Eventually, for all VLs with $A_{v} \gtrsim 1.0$, the system
reaches a {\it coupled\/} channel flow phase for large enough values
of ${\rm f}_{d}$.  The boundary of this phase is identified as the
current at which $P_{6}$ reaches a value of 1.  
At the transition $A_{v} \sim 0.75$ below which the coupled channel
phase never appears, the {\it greatest\/} maximum
noise power $S_{\rm max}$ is observed, indicating that the noise power
could be used as an experimental probe of this transition. 
At $A_{v} \sim 0.75$, 
vortex-vortex and vortex-pin interactions are nearly balanced, 
allowing the system to sample the largest number of metastable states.
The boundaries shown on the phase diagram mark crossover points rather
than sharp phase transitions.

In the coupled channel regime, we find a washboard frequency \cite{Harris}
in the voltage noise signal, shown in Fig.~\ref{fig:phase}(B), 
corresponding to the time
required for a vortex to move one lattice constant.
The magnitude of this washboard signal decreases when the system size
is increased,
indicating that the signal appears only when the region
of the VL sampled is locked into a single domain.
Thus experimentally it would be necessary to probe the voltage over a
very small area of a sample to observe a washboard frequency, such as
with local Hall probes.

In conclusion, we have obtained a new vortex dynamic phase diagram as
a function of lattice softness and driving current.
At high driving currents two distinct phases of flux flow appear:
in soft lattices, uncoupled channels with a smectic structure,
and in stiff lattices, coupled channels with crystalline-like order.
A signature of the crossover is observed in the voltage
noise, which is largest at the transition between the two phases.
In the coupled channel phase, a washboard frequency appears in the
voltage noise spectrum for small sample sizes.
Our results are in agreement with experiments.

We acknowledge helpful discussions with F. Pardo and S. Bhattacharya, 
and help from the UM CPC,
funded by NSF Grant No. CDA-92-14296.  CO was supported by NASA.

\end{document}